\documentclass{aa}
\usepackage{graphicx}

\begin{document}

\title{Radio observations of HD\,93129A: The earliest O star with the
       highest mass loss\,?}
\author{Paula Benaglia\inst{1,2}\thanks{Member of
           Carrera del Investigador, CONICET},
        B\"arbel Koribalski\inst{3}}

\institute{Instituto Argentino de Radioastronom\'{\i}a,
   C.C.5, (1894) Villa Elisa,  Argentina \and
    Facultad de Cs. Astron\'omicas y Geof\'{\i}sicas, UNLP,
    Paseo del Bosque S/N, 1900, La Plata, Argentina \and
   Australia Telescope National Facility, CSIRO, PO Box 76,
    Epping, NSW 1710, Australia}

\offprints{P. Benaglia \\ \email{paula@lilen.fcaglp.unlp.edu.ar}}

\date{Received July 31; accepted November 6, 2003}

\titlerunning{The HD\,93129A field}
\authorrunning{Paula Benaglia \& B\"arbel Koribalski}

\abstract{We present the results of radio continuum observations
towards the open cluster Tr\,14, where our main targets are the
early-type O stars HD\,93129A/B and HD\,93128. The observations
were carried out at 3\,cm (8.64 GHz) and 6\,cm (4.80 GHz) with the
Australia Telescope Compact Array. Only HD\,93129A (type O2 If*)
was detected; we measure flux densities of $S_{\rm 3cm} = 2.0 \pm
0.2$ mJy and $S_{\rm 6cm} = 4.1 \pm 0.4$ mJy. The resulting spectral
index of $\alpha = -1.2 \pm 0.3$ ($S_\nu \propto \nu^\alpha$)
indicates predominantly non-thermal emission, suggesting HD\,93129A
may be a binary system. We propose that the observed 3\,cm radio
emission is mostly coming from the non-thermal wind collision region
of a binary, and, to a lesser extent, from the thermal winds of the
primary and secondary stars in HD\,93129A. At a stellar distance of
2.8 kpc, we derive a mass-loss rate $\dot{M} = 5.1 \times 10^{-5}$
M$_{\odot}$\,yr$^{-1}$, assuming the thermal fraction of the 3\,cm
emission is $\sim$0.5.
 }

\maketitle

\bigskip

\section{Introduction}
Massive, early-type O stars at the upper end of the stellar main
sequence are known to have strong stellar winds, with velocities
up to $\sim$3000 km\,s$^{-1}$ (Prinja et al. 1990). Their
evolution into Wolf-Rayet (WR) stars is marked by high mass loss
rates of $10^{-7}$ to $10^{-5}$ M$_{\odot}$\,yr$^{-1}$ (e.g.,
Lamers \& Leitherer 1993). The transition can be described by
three Of phases (O -- O((f))  -- O(f) -- Of -- WR), where ``f''
stands for the presence of certain N\,{\sc iii} and He\,{\sc ii}
spectral lines (see Ma\'{\i}z Apell\'aniz \& Walborn 2002 for
details; also Table~1). Once the  H-rich outer layers of the star
are lost, the high temperature core is exposed, and the evolution
from Of to WR star is complete. The action of the stellar wind on
the ambient interstellar medium creates a cavity of hot, rarefied
gas, surrounded by a slowly expanding envelope (Benaglia \& Cappa
1999).

Wolf-Rayet stars have anomalously strong and broad emission lines,
which are thought to come from the rapidly expanding stellar wind,
but few absorption lines. With an average mass of 10 M$_{\odot}$,
their mass loss rate can reach values up to 10$^{-4}$
M$_{\odot}$\,yr$^{-1}$. Van der Hucht et al. (2001, 2002)
cataloged well over 200 WR stars in our Galaxy, more than half of
which belong to the WN class, i.e. they show strong helium and
nitrogen emission lines. A large fraction of the cataloged WR
stars ($\sim$40\%) are known to be binaries.

According to Crowther et al. (1995), the most massive ($>$ 60
M$_{\odot}$) O stars evolve from the Of stage directly into highly
luminous late WN(+abs) and WN7 stars without passing through an
intermediate Luminous Blue Variable or red supergiant phase. The
slightly less massive O stars evolve through either of those
intermediate phases into slightly less luminous WN8 stars (see
also Langer et al. 1994). The quantitative spectroscopic
properties show a smooth progression from Of stars to lower
excitation or late WN stars (WNL) (Bohannan 1990). Walborn et al.
(1992), Lamers \& Leitherer (1993) among others, suggested that
transition types Ofpe/WNL, and WNL stars, form a sequence of
increasing wind density.

Examples of transition objects are Sk--67$^0$ 22 (type
O3If*/WN6-A, Walborn 1982), Sk--71$^0$ 34 (type O4f/WN3, Conti \&
Garmany 1983), and, lately, the ones found in 30 Doradus (Massey
\& Hunter 1998). The mass loss rates of the transition objects are
expected to lie between those for Of and WR stars.

Almost 400 Galactic O stars have recently been compiled in the new
`Galactic O Star Catalogue' (GOS) by Ma\'{\i}z-Apell\'aniz \&
Walborn (2002). There are only about a dozen stars with spectral
types O3.5 or earlier; five of these belong to the clusters
Trumpler\,14 and Trumpler\,16, in the Carina region. Among them
one star, HD\,93129A, stands out as the earliest, most luminous,
and most massive star known in our Galaxy. Walborn et al. (2002)
re-analyzed spectra of HD\,93129A, previously classified as O3,
and find that it is the prototype of a new subcategory and the
sole representative of type O2 If* (see Table~1). HD\,93129A
appears to have the most extreme stellar winds regarding
ionization and terminal velocities (Walborn et al. 2002):
$v_\infty = 3200 \pm 200$ km\,s$^{-1}$, an effective temperature
$T_{\rm eff} = 52,000 \pm 1000$ K, and a stellar mass of $130 \pm
15$ M$_\odot$ (Taresch et al. 1997). Only $2.6"$ from HD\,93129A,
Walborn (1973) discovered a second star, HD\,93129B, of type O3.5
V((f+)). Another star of the same type, HD\,93128, is found at a
projected distance of $\sim20"$. The stellar parameters for all
three stars are listed in Table~1.

In the radio regime, free-free thermal emission from the ionized
wind of early-type OB and Wolf-Rayet stars can be modeled (Wright
\& Barlow 1975, Panagia \& Felli 1975) yielding a spectral index
($\alpha$) of typically $\sim 0.6 - 0.7$. Generally, the radio
flux densities are of the order of mJy for stars at distances up
to 2 kpc. About 80 WR stars (Abbott et al. 1986, Chapman et al.
1999) and $\sim$120 O stars (Bieging et al. 1989, Scuderi et al.
1998, Benaglia et al. 2001b, etc.) have been observed. Roughly
half of them were detected at least at one frequency. When
spectral indices could be measured, $\sim$40\% showed non-thermal
or composite indices.

In recent years the hypothesis was proposed (Chapman et al. 1999,
Dougherty \& Williams 2000) that non-thermal emission could be
indicative of binarity. In that case, part of the emission comes
from the colliding wind region (CWR) of two early massive stars.
Chapman et al. (1999) observed 36 WR stars with the ATCA at 3, 6,
13 and 20\,cm, obtaining an angular resolution of $\sim 1"$ at 3\,cm.
Dougherty \& Williams (2000) discussed the spectral index behaviour
of 25 WR stars, by taking into account observations carried out
mainly with the VLA, ATCA, MERLIN and the WSRT. Benaglia \& Romero
(2003) presented a table with the characteristics of the non-thermal
or composite WR emitters.

Unfortunately, in some particular cases where the non-thermal
emission is strong, not enough spectroscopic data are currently
available to investigate the binary status. For example,
CD--47$^{\circ}$ 4551 (CPD--47$^{\circ}$ 2963), an O4 III(f) star
(Walborn 1982), was observed with the ATCA (Benaglia et al. 2001b)
and detected with flux densities of $1.77 \pm 0.05$ mJy at 3\,cm
and $2.98 \pm 0.05$ mJy at 6\,cm ($\alpha = -0.89\pm0.06$). If the
star is a binary with another WR or OB star, the non-thermal
emission could be produced at the CWR of both stars.

Non-thermal radio emission can also be produced by the stellar
wind of a single star, as described by White (1985). In his model,
electrons are accelerated to relativistic energies by shocks in
the wind near the star, and emit radio radiation.
For strong shocks, the expected spectral index is +0.5. %
Though this is similar to the thermal index value, the non-thermal
flux density coming from a single stellar wind will suffer a break
at a frequency of a few GHz, and behave as $S(\nu) \propto
\log(\nu)$ for larger frequencies, while the thermal flux will
obey a $\alpha = 0.6 - 0.7$ law up to the infrared range.

Radio observations are very useful to derive the mass loss rates
of early-type stars (Leitherer et al. 1995). The source HD\,93129A
represents one of the most interesting candidates because it is the
only known example of the hottest supergiant class. With the help of
high resolution radio data, properties of the wind can be unveiled. In
detecting the stellar wind, we intend to identify its radio spectrum,
look for a possible colliding wind region, and compare the mass loss
rate with those derived from other methods. The observations will also
allow us the study  of other early-type stars in the region.

The paper is organized as follows: Section~2 summarizes the
characteristics of the target sources relevant to this study.
Section~3 describes the observations and data reduction. In
Section~4 we present the results, followed by the discussion in
Section~5. Section~6 contains our conclusions.

\section{The target stars in Trumpler\,14}
\begin{table*} 
\caption[h]{Stellar parameters adopted for the target stars}
\begin{flushleft}
\begin{tabular}{lrrrl}
 \hline
                    & HD\,93129A    & HD\,93129B    & HD\,93128 & \\
 \hline
$\alpha, \delta$ (J2000)
                    & 10 43 57.462$^{\rm (a)}$
                    & 10 43 57.638$^{\rm (b)}$
                    & 10 43 54.372$^{\rm (c)}$
                    & $^{\rm h\,m\,s}$ \\
                    &--59 32 51.27\,\,\,\,\,\,
                    &--59 32 53.50\,\,\,\,\,\,
                    &--59 32 57.37\,\,\,\,\,\,
                    & $^\circ\,'\,"$ \\
Spectral Class.     & O2 If*$^{\rm (d)}$
                    & O3.5 V((f+))$^{\rm (d)}$
                    & O3.5 V((f+))$^{\rm (d)}$     & \\
$v_\infty$          & $3200 \pm 200^{\rm (e)}$
                    & 3070$^{\rm (f)}$
                    & 3070$^{\rm (f)}$ & km\,s$^{-1}$ \\
$T_{\rm eff}$       & $52000 \pm 1000^{\rm (e)}$
                    & $52000 \pm 1500^{\rm (g)}$
                    & $52000 \pm 1500^{\rm (h)}$            & K \\
log ($L$/L$_\odot$) & $6.4 \pm 0.1^{\rm (e)}$
                    & 5.6$^{\rm (g)}$
                    & 5.7$^{\rm (g)}$   & \\
\hline
\end{tabular}
\end{flushleft}
References: (a): Hog et al. 2000;
        (b): Worley \& Douglass 1997;
        (c): see Ma\'{\i}z Apell\'aniz \& Walborn 2002;
        (d): Walborn et al. 2002
(f* means the N\,{\sc iv} 4058\AA\ emission lines are stronger than the
 N\,{\sc iii} 4640\AA\ emission lines, and HeII 4686\AA\ is observed in
 emission, while ((f+)) means weak N\,{\sc iii} 4634-4640-4642\AA\ emission
 lines, strong He\,{\sc ii} 4686\AA\ absorption lines, plus Si\,{\sc iv}
 4089--4116\AA\ emission lines);
        (e): Taresch et al. 1997;
        (f): interpolated from Prinja et al. 1990;
            (g): see text;
        (h): Vacca et al. 1996.
\label{table1}
\end{table*}

The field of view of the present data (see Section~3) corresponds
to the cluster Trumpler\,14. Lists of the catalogued members can
be found in Massey \& Johnson (1993), and recently
DeGioia-Eastwood et al. (2001). A wide range of spectroscopic or
photometric distances can be found for Tr\,14 in the literature.
Humphreys \& McElroy (1984) derived a distance of 3.47 kpc.
Morrell et al. (1988) discussed two possible distances, 2.8
(variable $R$, from Tapia et al. 1988) and 3.45 kpc ($R$ = 3.2),
depending on the value taken for the ratio of total to selective
absorption $R$. Massey \& Johnson (1993) and V\'azquez et al.
(1996) found similar values of 3.2 and 3.1 kpc, respectively. Most
recently, Tapia et al. (2003) computed spectroscopic parallaxes to
derive a distance of 2.8 kpc. We adopt the latter value throughout
this paper. At this distance, the projected cluster size of $140"$
(Tapia et al. 2003) corresponds to 1.9 pc. Chandra observations
towards the $\eta$ Carina region (Evans et al. 2003) revealed
strong X-ray emission from the stars HD\,93129A/B. Their combined
X-ray luminosity is $L_{\rm X}(0.5-2.4{\rm keV}) = 2.9 \times
10^{33}$ erg\,s$^{-1}$, adjusted to a distance of 2.8 kpc.

Penny et al. (1993) presented radial velocity measurements for
seven stars in Tr\,14, with mean values of $-4.4 \pm 7.4$
km\,s$^{-1}$ for HD\,93129A, $7.2 \pm 2.4$ km\,s$^{-1}$ for
HD\,93129B and $7.3 \pm 3.3$ km\,s$^{-1}$ for HD\,93128. They
concluded that none of them is a strong binary candidate (see also
Conti et al. 1977, 1979). Based on the standard deviation in the
mean stellar radial velocities, they estimated that the typical
random motions in this cluster are 8.5 km\,s$^{-1}$. For
HD\,93129A Penny et al. placed an upper limit of $\sim$10
km\,s$^{-1}$ on the semiamplitude of any short-period binary. To
search for possible long-period binaries they obtained speckle
observations, with negative results. Surprisingly, recent HST
observations indicate that HD\,93129A is a binary (Walborn 2002)
with a separation of 55 mas (Walborn 2003, priv. comm.; Nelan et
al., in prep.), which could have been detected by Penny et al.
(1993).

The optical positions of the three earliest stars in the observed
field were taken from the GOS catalog, and come from three
different sources, varying in precision: the Tycho-2 catalog (Hog
et al. 2000) lists a very accurate position ($\sigma \sim 5$ mas)
for HD\,93129A, whereas the position of HD\,93129B originates from
the Washington Double Star catalog (Worley \& Douglass 1997) and
that of HD\,93128 was measured from NTT data, according to
Ma\'{\i}z-Apell\'aniz \& Walborn (2002). For a summary of the
stellar parameters see Table~1.

\paragraph{\bf HD\,93129A.} The spectrum of this early O2 If* star is
very similar to Wolf-Rayet stars of spectral type WN6 or WN7 (Walborn
et al. 2002). Taresch et al. (1997) studied the spectrum from
optical to FUV range, concluding that its strong wind is radiatively
driven. By fitting an H$\alpha$ profile they obtained a mass loss rate
of $1.8 \times 10^{-5}$ M$_{\odot}$\,yr$^{-1}$ compared to $2.2 \times
10^{-5}$ M$_{\odot}$\,yr$^{-1}$ derived from H$\alpha$ measurements by
Puls et al. (1996; assuming $T_{\rm eff}$ = 50,500 K).

\paragraph{\bf HD\,93129B and HD\,93128.}
The effective temperature of HD\,93128 is $52000 \pm 1500$ K
(Vacca et al. 1996). For HD\,93129B, which has the same spectral
type as HD\,93128, we adopt the same value. The stellar
luminosities were derived taking the absolute visual magnitudes
from Walborn et al. (2002).

\paragraph{\bf HD\,93250.} This star is a member of the Tr\,16 cluster,
located $\sim10'$ East of HD\,93129A/B (well outside our primary
beam). Leitherer et al. (1995) reported a 3\,cm flux density of
$1.36 \pm 0.17$ mJy and a 3$\sigma$ upper limit of 3.57 mJy at
6\,cm. Assuming a purely thermal spectrum, a distance of 2.2 kpc,
a wind velocity of $v_\infty$ = 3200 km\,s$^{-1}$ and an effective
temperature of $T_{\rm eff}$ = 50,000\,K ($T_{\rm e}$ = 0.4
$T_{\rm eff}$ = 20,000\,K according to Drew 1989) Leitherer et al.
derived a mass loss rate of $4.1 \times 10^{-5}$
M$_{\odot}$\,yr$^{-1}$. Considering a re-classified spectral type
of O3.5 V((f+)) (Walborn et al. 2002) and a distance of 2.5 kpc to
Tr\,16 (Tapia et al. 2003) we revise the mass loss rate to $5
\times 10^{-5}$ M$_{\odot}$\,yr$^{-1}$ (see Section~4.1).

\section{Observations and Data Reduction}
The observations presented here were obtained with the Australia
Telescope Compact Array (ATCA) in January 2003 using the 6B
configuration. The target field, which contains three massive Of
stars, was observed simultaneously at two frequencies: 8.64 GHz
(3\,cm) and 4.80 GHz (6\,cm), with a total bandwidth each of 128
MHz over 32 channels. The maximum baseline was 6\,km, resulting in
angular resolutions around $1"-2"$. The size of the primary beam
is $10'$ at 6\,cm and $5'$ at 3\,cm. The pointing position was
$\alpha,\delta$(J2000) = $10^{\rm h}\,43^{\rm m}\,56^{\rm s}$,
$-59^{\circ}\,32'\,51"$, slightly offset from the three stars
HD\,93129A/B and HD\,93128. The field was observed during
intervals of 8 to 15 min, depending on the atmospheric phase
stability, interleaved with short observations of the nearby phase
calibrator 1045--62, and spanning an LST range of 12\,h. The total
time on the target field was nearly 300 minutes.

The data were reduced in {\sc miriad} using standard procedures.
The flux density scale was calibrated with observations of the
primary calibrator PKS B1934--638, assuming flux densities of
2.84\,Jy at 3\,cm and 5.83\,Jy at 6\,cm (Reynolds 1994). After
data calibration the visibilities were Fourier-transformed using
`natural'- and `uniform'-weighting. To overcome confusion due to
diffuse emission from extended sources in or towards the observed
fields, the shortest baselines were removed. For the same reason
`uniform'-weighting resulted in the lowest noise levels: 0.11 and
0.21 mJy\,beam$^{-1}$ at 3 and 6\,cm, respectively. The
synthesized beam is $1.1" \times 1.0"$ at 3\,cm and $1.9" \times
1.7"$ at 6\,cm.

\section{Results}
Figs.~1 and 2 show the central part of the resulting 3 and 6\,cm
radio continuum images. The optical positions of the three target
stars (see Table~1) are marked.

Only one radio source, HD\,93129A, is detected. Using the {\sc miriad}
task {\em imfit} we fit the position and flux density (of a point
source) at both frequencies resulting in
$\alpha,\delta$(J2000) = $10^{\rm h}\,43^{\rm m}\,57^{\rm s}.45$,
$-59^{\circ}\,32'\,51.36"$, $2.0 \pm 0.2$ mJy at 3\,cm, and
$\alpha,\delta$(J2000) = $10^{\rm h}\,43^{\rm m}\,57^{\rm s}.47$,
$-59^{\circ}\,32'\,51.23"$, $4.1 \pm 0.4$ mJy at 6\,cm. From the 3
and 6\,cm flux densities we derive a spectral index of $\alpha =
-1.2 \pm 0.3$ ($S_\nu \propto \nu^\alpha$) indicating predominantly
non-thermal emission. The measured flux densities for HD\,93129A
and the 3$\sigma$ upper limits for HD\,93129B and HD\,93128 are
listed in Table~2.

The separation between the fitted 3 and 6\,cm radio continuum
positions for HD\,93129A is $0.2"$, slightly larger than the
positional uncertainty (which is approximately the size of the
synthesized beam divided by signal-to-noise ratio of the source).
The radio and optical positions agree within the given
uncertainties.

\begin{figure} 
\centering
 \includegraphics[width=7.5cm,angle=-90]{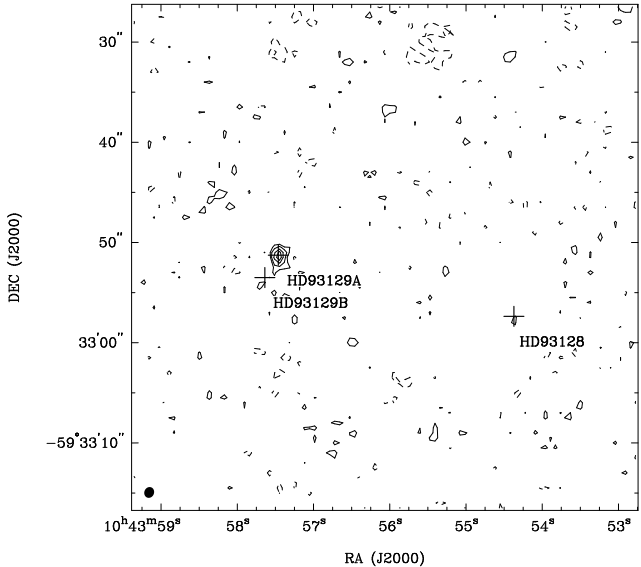}
  \caption{ATCA 3\,cm radio continuum image towards Tr\,14. The optical
     positions of three stars (see Table~1) are marked. The contour
     levels are --0.22, 0.22 ($2\sigma$), 0.44, 0.66, 0.88, and 1.10
     mJy\,beam$^{-1}$. The synthesized beam ($1.1" \times 1.0"$) is
     displayed at the bottom left.}
  \label{fig3cm} 
\end{figure}

\begin{figure} 
\centering
 \includegraphics[width=7.5cm,angle=-90]{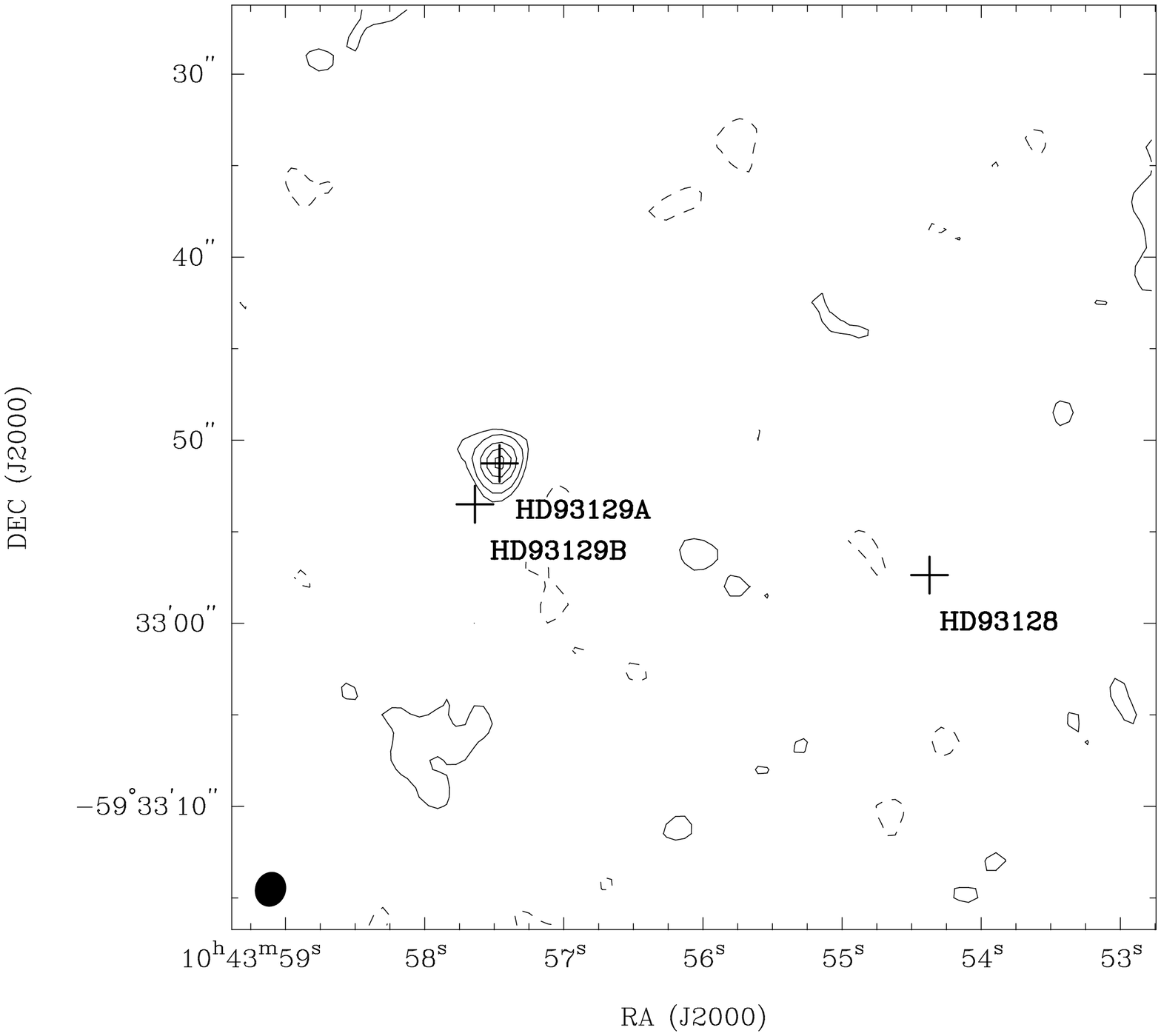}
  \caption{ATCA 6\,cm radio continuum image towards Tr\,14. The optical
     positions of three stars (see Table~1) are marked. The contour
     levels are --0.42, 0.42 ($2\sigma$), 0.84, 1.26, 1.68, and 2.10
     mJy\,beam$^{-1}$. The synthesized beam ($1.9" \times 1.7"$) is
     displayed at the bottom left.}
\label{fig6cm} 
\end{figure}

\begin{table} 
\caption{Flux densities (or 3$\sigma$ upper limits)
         of the target stars.}
\begin{center}
\begin{tabular}{lcc}
\hline
Star        & $S_{\rm 3cm}$ & $S_{\rm 6cm}$ \\
            & (mJy)         & (mJy)         \\
\hline
 HD\,93129A & $2.0\pm0.2$   & $4.1\pm0.4$   \\
 HD\,93129B & $<0.33$       & $<0.63$       \\
 HD\,93128  & $<0.33$       & $<0.63$       \\
\hline
\end{tabular}
\end{center}
\end{table}

The Simbad database lists 250 Galactic objects within a radius of
$2.5'$ around our pointing position (see Section~3), at least 11
of which are O-type stars, including the three target stars.
Table~3 lists the eight additional stars, none of which was
detected. The same $3\sigma$ upper limits apply.

\begin{table*} 
\caption{Additional O-type stars within the observed 3\,cm field}
\begin{flushleft}
\begin{tabular}{llllrr}
\hline Name      & \multicolumn{2}{c}{$\alpha, \delta$ (J2000)}
                 & Spectral type
                 & $\dot{M}_{\rm exp}$
                 & $S^{\rm T}_{\rm exp,3cm}$  \\
                 & (h m s) & ($^\circ\,'\,"$)
                 &
                 & M$_{\odot}$ yr$^{-1}$
                 & mJy                        \\
\hline
Cl* Trumpler 14 MJ 165 & 10 43 55.43 & --59 32 49.8
   & O8 V$^{\rm (a)}$         & $8.5 \times 10^{-7}$ & 0.010\\
CPD-58 2620            & 10 43 59.92 & --59 32 25.4
   & O6.5 V((f)) $^{\rm (b)}$ & $1.7 \times 10^{-6}$ & 0.016\\
Cl* Trumpler 14 MJ 150 & 10 43 53.75 & --59 33 29.0
   & O9: V$^{\rm (a)}$        & $4.5 \times 10^{-7}$ & 0.005\\
Cl* Trumpler 14 MJ 127 & 10 43 48.82 & --59 33 24.8
   & O9 V$^{\rm (a)}$         & $4.5 \times 10^{-7}$ & 0.005\\
CPD-58 2611            & 10 43 46.69 & --59 32 54.8
   & O6 V((f))$^{\rm (b)}$    & $2.1 \times 10^{-6}$ & 0.021\\
HD 93160            & 10 44 07.27 & --59 34 30.8
   & O6 III((f))$^{\rm (b)}$  & $4.9 \times 10^{-6}$ & 0.064\\
HD 93161AB             & 10 44 08.90 & --59 34 34.9
   & O6.5 V((f))$^{\rm (b)}$  & $1.7 \times 10^{-6}$ & 0.016\\
LS 1832                 & 10 44 06 & --59 35 09.0
   & O6$^{\rm (c)}$           & $2.1 \times 10^{-6}$ & 0.021\\
\hline
\end{tabular}
\end{flushleft}
References: (a): Massey \& Johnson 1993;
            (b): Ma\'{\i}z-Apell\'aniz \& Walborn 2002;
            (c): Reed \& Beatty 1995.
\end{table*}

\subsection{Mass loss rates}
According to Wright \& Barlow (1975), the stellar mass loss rate,
$\dot{M}$, of an optically thick ionized wind can be expressed in
terms of the measured thermal flux density, $S_\nu$, at a frequency
$\nu$, from radio to IR ranges, as:

\begin{equation} 
 \dot{M} = 5.32 \times 10^{-4} \, \frac{S_\nu^{3/4}\,d^{3/2} \,
           v_\infty \, \mu}{Z\, \sqrt{\gamma\, g_\nu \,\nu}} \,\,\,\,\,\,
           {\rm M}_{\odot} {\rm yr}^{-1}~.
\end{equation}

Here $d$ is the stellar distance in kpc, $v_\infty$ the wind
terminal velocity in km\,s$^{-1}$,  $S_\nu$ the thermal flux
density in mJy, and $\nu$ the frequency in Hz; $\mu$ stands for
the mean molecular weight of the ions, $\gamma$ is the mean number
of electrons per ion and $Z$ the r.m.s. ionic charge. For a fully
ionized H-He gas, $\mu = 1.38$. In a first approximation, $Z
\approx \gamma \sim 1$. We will take those values for all
calculations throughout the text. The Gaunt factor $g_\nu$ can be
approximated by:

\begin{eqnarray} 
 g_\nu = 9.77\, \left( 1 + 0.13 \log
         \frac{(0.4\times T_{\rm eff})^{3/2}}{Z\nu} \right)~.
\end{eqnarray}

From the derived spectral index of $\alpha = -1.2 \pm 0.3$ we know
that the observed radio continuum emission consists of thermal and
non-thermal components. Their relative contributions (and
locations) are unknown since the detected source is essentially
unresolved. If $f_{\rm T}$ is the fraction of the thermal emission
at 3\,cm with $S^{\rm T}_{\rm 3cm} = f_{\rm T} \times S_{\rm
3cm}$, $f_{\rm T} = 0 - 1$, then the mass loss rate of HD\,93129A
is $\dot{M} = (f_{\rm T})^{3/4} \times 8.6 \times 10^{-5}$
M$_\odot$\,yr$^{-1}$. For $f_{\rm T} \la 1$ this would be one of
the highest mass loss rates ever derived for a massive, early-type
star.  For HD\,93129B and HD\,93128 we derive 3$\sigma$ upper
limits of $\dot{M} < (f_{\rm T})^{3/4} \times 2.13 \times 10^{-5}$
M$_\odot$\,yr$^{-1}$.\\

Given a stellar mass, luminosity, effective temperature, and wind terminal
velocity, the models of Vink et al. (2000) can be used to compute a
theoretical --- or expected --- mass loss rate, $\dot{M}_{\rm exp}$. Using
Vink's program (see astro.imperial.ac.uk/$\sim$jvink/) and the values
given in Table~1, we calculate $\dot{M}_{\rm exp} = 2.12 \times 10^{-5}$
M$_\odot$\,yr$^{-1}$ for HD\,93129A. While this value is in good agreement
with the H$\alpha$-derived mass loss rates (Puls et al. 1996, Taresch et
al. 1997), it is a factor four smaller than that derived from the measured
3\,cm radio continuum flux density, suggesting $f_{\rm T} = 0.15$.
For HD\,93129B and HD\,93128 we calculate $\dot{M}_{\rm exp}$ of
$0.3 \times 10^{-5}$ M$_\odot$\,yr$^{-1}$.

Table~3 lists the expected stellar mass loss rates for the other
eight O stars in the field, estimated as for the earliest stars
and assuming a distance of  2.8 kpc. The luminosities, masses, and
effective temperatures were taken from Vacca et al. (1996). The
terminal velocities were interpolated from Table 3 of Prinja et
al. (1990) according to spectral type. The expected thermal flux
density at 3\,cm was computed inverting Eq.~(1), and is listed in
the last column.

\section{Discussion}
\subsection{A CWR between HD\,93129A and B\,?}
 The distance of $D = 2.6"$ = 7300 AU between coordinates of
HD\,93129A and B is a lower limit to their absolute separation. If
a colliding wind region exists between the two stars, it would be
located at projected distances $r_{\rm A}$, $r_{\rm B}$ (see
Eichler \& Usov, 1993):

\begin{equation}
r_{\rm A}=\frac{1}{1+\eta^{1/2}}D~,\;\;\;\;\;
r_{\rm B}=\frac{\eta^{1/2}}{1+\eta^{1/2}}D~.
\end{equation}

The parameter $\eta$ is defined in terms of the wind terminal velocities
and the stellar mass loss rates $\dot{M}$:
\begin{equation}
\eta=\frac{\dot{M_{\rm B}} \times v_{\infty,{\rm B}}}{\dot{M_{\rm A}}
      \times v_{\infty,{\rm A}}}~.
\label{eta}
\end{equation}

Using the derived theoretical values for $\dot{M_{\rm A}}$ and
$\dot{M_{\rm B}}$, we find $\eta$ = 0.14. Thus $r_{\rm A} = 1.9"
\approx 5350$ AU, and $r_{\rm B} = 0.7" \approx 1950$ AU. The
putative CWR would be centered around $\alpha,\delta$(J2000)$_{\rm
CWR}$ = $10^{\rm h}\,43^{\rm m}\,57.59^{\rm s}$,
$-59^{\circ}\,32'\,52.9"$, but no emission was detected with the
ATCA. The same 3$\sigma$ upper limits apply (see Table~2).

Assuming a spectral index of $\alpha = -0.5$ (Eichler \& Usov
1993), we derive an upper limit to the non-thermal radio
luminosity from a CWR between stars A and B of $L_{\rm obs} < 1.4
\times 10^{28} {\rm ~erg\,s}^{-1}$. The expected non-thermal
luminosity from a possible A/B binary is difficult to estimate due
to the poor knowledge of system parameters such as size of the CWR
and the magnetic field. Moreover, according to Wright \& Barlow
(1975), the radii of the radio photospheres at 3 and 6\,cm would
be $\sim$30 and 60 AU, ($<< r_{\rm A}$, $r_{\rm B}$),
respectively, if $\dot{M} = 5 \times 10^{-5}$
M$_\odot$\,yr$^{-1}$. The existence of a CWR between HD\,93129A
and B is very unlikely (see Section~5.4).

\subsection{A CWR between the binary components of HD\,93129A\,?}
Walborn (2002) recently reported that HD\,93129A is a binary
system. Using the HST Fine Guidance Sensor Interferometer, Nelan
et al. (2003, in prep.) resolved it into a 55 mas binary with a
magnitude difference of 0.9 mag (Walborn 2003, priv. comm.). In
the binary system, we will call ``a'' the O2 primary, and ``b'',
the secondary. The total $V$ magnitude measured for HD\,93129A is
7.2 mag (Hog et al. 1998). Given that $V_{\rm a} = 7.6$ mag, and
$V_{\rm b} = 8.5$ mag (Walborn 2003, priv. comm.), the secondary
has a magnitude similar to that of HD\,93129B ($V=8.9$ mag) and
HD\,93128 ($V=8.8$ mag) (Walborn et al. 2002). The  four stars are
probable coeval, and their spectral types should be similar (E.
Nelan et al., in preparation).

The binary separation (55 mas = 154 AU) implies that the stellar
winds are likely to interact. In fact, there is a high probability
that the system is an O+O colliding wind binary. Additional
support to this idea can be found by studying the radio spectral
index (see Section~5.3).  Assuming that the companion is an O3.5 V
star with the same terminal velocity, effective temperature,
luminosity, and expected mass loss rate than HD\,93129B (see
Table~1), we find $\eta = 0.14$, $r_{\rm Aa}= 112$ AU, and $r_{\rm
Ab} = 42$ AU. The binary separation of HD\,93129A is similar to
that of WR\,146 (210 AU, Setia Gunawan et al. 2000), thus it is
reasonable to assume a similar extent for the CWR, i.e. $\sim$50
AU. We estimate the equipartition magnetic field for the CWR
according to Miley (1980) and find $B_{\rm CWR} \approx 4$ mG.
Using the magnetic field structure given by Eichler \& Usov
(1993), the surface magnetic field of HD\,93129Aa can be
extrapolated to $B_* \sim 100$ G, if the stellar rotational
velocity is $0.1 v_\infty$ (Conti \& Ebbets 1997). This estimate
is consistent with the few observational values available (Donati
et al. 2001, 2002).

Substituting the parameters of HD\,93129Aa and Ab into
Eqs. (15) and (16) from Eichler \& Usov (1993) we obtain:
\begin{eqnarray}
  L_{\rm rad} \approx
  &7.4 \times 10^{31}\,\beta
  \, \left(\frac{\epsilon}{10^{-2}} \right)
  \, \left(\frac{\xi}{10^{-3}} \right)
  {\rm erg\,s}^{-1}
\end{eqnarray}
for the expected non-thermal radio luminosity of the CWR, and
$\xi \approx 5 \times 10^{-5}$.
Taking typical values of $\beta \sim 0.1$,
 $\epsilon \sim 10^{-2}$, we calculate $L_{\rm rad} \approx
 4 \times 10^{29} {\rm ~erg\,s}^{-1}$.

From the observed flux densities we obtain a radio luminosity of
$L_{\rm obs} = 1.12 \times 10^{29}$ erg\,s$^{-1}$ for HD\,93129A
between 4.8 and 8.64 GHz, which is in good agreement with the total
expected one.

In a colliding wind binary system like the one described here, a
total X-ray luminosity due to Bremsstrahlung can be computed.
Following Usov (1992) we obtain $2.2 \times 10^{33}$
erg\,s$^{-1}$, very close to the observed X-ray luminosity by
Evans et al. (2003).

\subsection{Spectral index considerations}
The spectral index of $-1.2\pm0.3$ derived from the observations
presented here is much steeper than that estimated by White
(1985), $\alpha = +0.5$, for synchrotron emission from a single
star. Steeper indices can result in the CW scenario as the effect
of the softening of the particle spectrum due to inverse Compton
(IC) losses (Benaglia \& Romero 2003). At a CWR the accelerated
electrons follow a power-law spectrum $N(E) \propto E^{p}$, with
$p=-2$. Both synchrotron and IC processes help electrons to lose
energy. At high energies, the radiative cooling would be dominated
by IC losses, due to the strong photon fields at the CWR. At the
energy $E_{\rm b}$ at which the cooling and escape times are equal
(e.g. Longair 1997, p. 281), IC losses produce a break in the
spectrum, from an index $p$ to $p'= p-1$. The same population of
relativistic electrons is capable of losing energy via synchrotron
emission, producing a flux $S(\nu) \propto \nu^\alpha$, where
$\alpha = (p+1)/2$. Thus, for $E < E_{\rm b}$, $\alpha \approx
-0.5$, and for $E > E_{\rm b}$, $\alpha \approx -1$. This last
value seems to be consistent with the spectral index derived from
the present observations. The break energy can be computed if the
following set of parameters is known (see Benaglia \& Romero
2003): the mass loss rates, wind terminal velocities, and
luminosities of both components, and the magnetic field at the
CWR, together with the size $s$ of the non-thermal source.
Replacing the values adopted or obtained above, the break
frequency in the radio range will be a few GHz. ATCA observations
at 1.4 and 2.4 GHz are necessary to test this scenario.

The relative importance of the processes already mentioned can be
measured by estimating the time needed for an electron to lose all
its energy, either in the form of synchrotron radiation ($t_{\rm
syn}$) or IC scattering ($t_{\rm IC}$). The importance of
synchrotron and inverse Compton radiation for electrons of the
same energy can be measured computing the ratio between the
corresponding lifetimes (Longair 1997, p.274):

\begin{equation}
\frac{t_{\rm syn}} { t_{\rm IC} }= \frac{U_{\rm rad}}{U_{\rm
mag}},~\label{ratio-ts}
\end{equation}

\noindent where $U_{\rm rad}$ and $U_{\rm mag}$ are the energy
densities of the radiation field and of the magnetic field
respectively. Using the observed radio luminosity $L_{\rm obs}$, we get
$U_{\rm rad} = 2.5 \times 10^{-9}$ eV m$^{-3}$.
The variable  $U_{\rm mag}$ is $6.4 \times 10^{-15}$ eV m$^{-3}$.
The corresponding time ratio is about $\sim4 \times 10^5$, thus
$t_{\rm syn} >> t_{\rm IC}$. Following this last estimate, and
under the conditions mentioned above, IC losses should dominate
over synchrotron ones.

\subsection{Colliding wind emission from other binaries}
There are only a few examples (WR~146, WR~147, Cyg OB2 No. 5) of
early-type binary systems in the literature where radio emission
from the CWR has been observed. WR~146 is a WC6+O8 binary (van der
Hucht 2001). Dougherty et al. (2000) observed the stars with
MERLIN at 5 GHz twice, and with the VLA at 1.4, 5, 8.5 and 22 GHz.
The combination of their MERLIN data with the 22 GHz VLA data
allowed  them to detect two thermal sources, coincident with the
WR and OB stars, plus a non-thermal one in-between, nearer to the
OB star, and possibly coming from the CWR. At a distance of 1.25
kpc (van der Hucht et al. 2001), the separation of the stellar
components is $\sim$210 AU, and the size of the non-thermal source
is $\sim$50 AU (Setia Gunawan et al. 2000). The flux densities
quoted in the literature are: $31.4\pm0.4$ mJy and $28.5\pm0.3$
mJy at 5 GHz for the source identified as the CWR, $7.0\pm1.3$ mJy
at 22 GHz for the emission possibly related to the WR star, and
$10.4\pm1.0$ mJy from the OB star. The ratios of the non-thermal
to thermal flux densities are $\sim$3 for the O star, and $\sim$4
for the WR star.

Contreras \& Rodr\'{\i}guez (1999) presented 8.5 GHz-VLA
observations of the star WR~147, a binary system formed by WN8
(h)+B0.5 V components (van der Hucht et al. 2001), at a distance of
650 pc (van der Hucht 2001). They detected a southern, thermal radio
source coincident with the WR component, and a northern,
non-thermal source. The latter is interpreted as originating at the
interaction zone between the winds of the WR and OB stars. Their
averaged measured flux densities are $28.4\pm0.5$ mJy from the
thermal source, and $10.4\pm0.5$ mJy from the non-thermal source.
The stellar separation is
$\sim$420 AU (Setia Gunawan et al. 2001). The non-thermal to thermal
flux density ratio is $\sim$0.4.

Cyg OB2 No. 5 seems to be formed by three stars: an O7 Ia+Ofpe/WN9
contact binary, and a B0 V star in a larger orbit (Contreras et
al. 1997, Rauw et al. 1999), located at 1.8
kpc (e.g. Waldron et al. 1998). The separation between the contact
pair and the third component is about 1700 AU. Radio observations
by Persi et al. (1990) and Miralles et al. (1994) at 1.4, 5, 8.5 and 15
GHz, showed that the source coincident with the contact pair
presented flux variations, defining a ``high'' emission state
($S_{\rm 8.5GHz}$ = 5 -- 7 mJy, and $\alpha \sim 0$), and a
``low'' emission state ($S_{\rm 8.5GHz}$ = 1 --2 mJy, and $\alpha
\sim 0.6$). Contreras et al. (1996, 1997) performed VLA
observations at 5 and 8.5 GHz, detecting a non-thermal source
($\alpha=-2.4\pm0.6$) near the B0 V star, and proposed  it was a
CWR in the system. They measured a flux density at 8.5 GHz of
$0.31\pm0.02$ mJy. The ratio of the emission from the weaker
non-thermal source to the emission of the contact pair at the
lower state, is about 0.2.

The system HD\,93129A presents a binary separation similar to that
of WR\,146.  The ratio of non-thermal to thermal emission is
unknown. If the magnetic field at the CWR is equal to the
equipartition field derived above, $\sim$4 mG, then it is similar
to those derived for WR\,147 (5 mG) and WR 146 (25 mG); see
Benaglia \& Romero (2003). The scarcity of the examples and the
huge differences in the ratios of CWR-non-thermal to
stellar-thermal flux densities measured for the three cases from
the literature, make the results of the comparison inconclusive.

\subsection{The radio spectrum of HD\,93129A}
To separate the thermal and non-thermal contributions to the
spectrum of HD\,93129A, further radio observations are needed at
lower and, particularly, at higher radio frequencies. These can
be obtained with the ATCA at 1.4, 2.3, 16--25, and 85--91 GHz.
The system can be resolved with the Long Baseline Array, but
the flux densities of the thermal winds are likely too low for
a detection, and an absolute position for the CWR would be
difficult to obtain.

\begin{figure} 
\centering
\includegraphics[width=8cm]{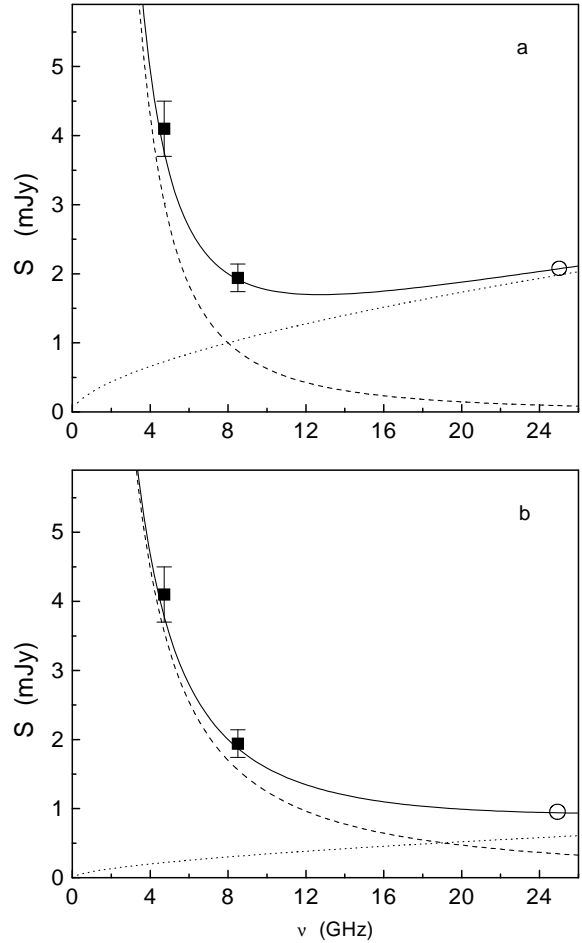}
  \caption{a) Measured and predicted radio continuum flux densities
   for HD\,93129A in the frequency range from $\sim$1 to 25 GHz. The
   two measurements at 4.8 and 8.64 GHz are indicated by black squares.
   Assuming that the flux density measured at 8.64 GHz, $S_{\rm 3cm}$,
   consists to equal parts of thermal and non-thermal emission
   ($f_{\rm T}$ = 0.5), and that the thermal emission has a spectral
   index of $\alpha_{\rm T}$ = 0.6, we can fit the observed flux densities
   by adding a non-thermal component  with $\alpha_{\rm NT} = -2.1 \pm 0.3$.
   The dotted and dashed lines show the thermal and non-thermal radio
   emission, respectively. The solid line represents the combined
   thermal and non-thermal radio emission for this model. b) As Fig.~3a,
   but assuming $f_{\rm T}$ = 0.15. In this case, we
   can fit the observed flux densities by adding a non-thermal component
   with $\alpha_{\rm NT} = -1.4 \pm 0.3$. }
\end{figure}

If the actual mass loss rate of HD\,93129A is $\dot{M} = 2.1 \times
10^{-5}$ M$_\odot$ yr$^{-1}$, i.e. $f_{\rm T}$ = 0.15 (see Section~4.1),
the thermal and non-thermal contributions to the 3\,cm radio flux
density are $\sim$0.3 and $\sim$1.7 mJy, respectively. Assuming a
spectral index of $\alpha_{\rm T}$ = 0.6 for the thermal component,
we find the spectral index for the non-thermal component is
$\alpha_{\rm NT} \sim -1.4 \pm 0.3$. The expected total flux densities
(see Fig.~3a) are $\sim$1.2, $\sim1$, and $\sim$1.3 mJy at 16, 25 and
90 GHz, respectively.

In contrast, for $f_{\rm T}$ = 0.5 we find $\alpha_{\rm NT} \sim -2.1
\pm 0.3$, similar to that derived for other Of systems (e.g., Cyg OB2
No. 5, Contreras et al. 1996, 1997). At 16, 25 and 90 GHz we expect
$\sim$1.7, $\sim$2, and $\sim$4.1 mJy (see Fig.~3b). In this case, the
radio derived mass loss rate of the system is $5.1 \times 10^{-5}$
M$_\odot$ yr$^{-1}$. The two scenarios can be distinguished with
high-frequency ATCA synthesis observations.

\section{Conclusions}
The field of view towards the Galactic cluster Tr 14 has been observed
with the ATCA in the radio continuum at 4.80 GHz (3\,cm) and 8.64 GHz
(6\,cm). Only one source, the O2 If* star HD\,93129A, was detected with
flux densities of $2.0\pm0.2$ mJy at 3\,cm, and $4.1\pm0.4$ mJy at 6\,cm,
resulting in a very steep spectral index of $\alpha = -1.2\pm0.3$. This
star, recently reported as binary, is the earliest known O-star and
potentially the most massive system in the Galaxy, followed by $\eta$
Carina.
The observed radio emission is most likely coming from (a) the thermal
winds of both binary components and (b) the non-thermal emission from
the colliding wind region. The negative spectral index of HD\,93129A
suggests that the latter is the dominant emission mechanism.

We derive a mass loss rate of $(f_{\rm T})^{3/4} \times 8.6 \times
10^{-5}$ M$_\odot$ yr$^{-1}$ for HD\,93129A and 3$\sigma$ upper limits
of $(f_{\rm T})^{3/4} \times 2.13 \times 10^{-5}$ M$_\odot$ yr$^{-1}$
for HD\,93129B and HD\,93128, where $f_{\rm T}$ is the thermal
fraction of the 3\,cm radio continuum emission.

Although the secondary star in the binary HD\,93129A remains yet
unclassified, the presence of intense non-thermal emission and
high radio luminosity together suggest it is another massive,
early-type star.

The lack of radio emission near HD\,93129B, and the extremely large separation
of HD\,93129A/B (7300 AU at 2.8 kpc) conspires against the existence
of a colliding wind region in-between the stars.

It is crucial to perform observations (i) at lower frequencies, where
the thermal contributions should be negligible, and an accurate
non-thermal index can be derived; this would help to investigate
the processes responsible for the radiation detected; and (ii) at
high frequencies, at which non-thermal emission decreases,
and the estimation of the thermal flux will enable us to derive the
value of the true mass loss rate of HD 93129A.

\section*{Acknowledgments}
We would like to thank the anonymous referee for comments and
suggestions that help to improve the presentation of
our investigation.
We are indebted to E. Nelan and N. A. Walborn for allowing us to share
their HST results before publishing, and
to N. A. Walborn and G. E. Romero for a critical
reading of the manuscript. P.B. is grateful to the ATNF staff at
Sydney and Narrabri. This research has been supported mainly by
Fundaci\'on Antorchas (P.B.), and has made use of the SIMBAD
database operated at CDS, Strasbourg, France.

\section*{References}
Abbott, D.C., Bieging, J.H., \& Churchwell, E. 1986, ApJ 303, 239\\
Benaglia, P., \& Cappa, C.E. 1999, A\&A 346, 979\\
Benaglia, P., Romero, G.E., Stevens, I.R., \& Torres, D.F. 2001a,
  A\&A 366, 605\\
Benaglia, P., Cappa, C.E., \& Koribalski, B.S. 2001b, A\&A 372, 952\\
Benaglia, P., \& Romero, G.E. 2003, A\&A 399, 1121\\
Bohannan, B. 1990, in ``Properties of Hot Luminous Stars'', ed. C.D.
  Garmany, Brigham Young Press, 39\\
Chapman, J.M., Leitherer, C., Koribalski, B.S., et al. 1999, ApJ 518, 890\\
Chlebowski, T., \& Garmany, C.D. 1991, ApJ 368, 241\\
Conti, P.S., Leep, E.M., \& Lorre, J.J. 1977, ApJ 214, 759\\
Conti, P.S., Niemela, V.S., \& Walborn, N.R. 1979, ApJ 228, 206\\
Conti, P.S., \& Garmany, C.D. 1983, PASP 95, 411\\
Conti, P.S., \& Ebbets, D. 1997, ApJ 213, 438 \\
Contreras, M.E., Rodr\'{\i}guez, L.F., G\'omez, Y., et al. 1996,
  ApJ 469, 329\\
Contreras, M.E., Rodr\'{\i}guez, L.F., Tapia, M., et al.
  1997, ApJ 488, L153\\
Contreras, M.E., \& Rodr\'{\i}guez, L.F. 1999, ApJ 515, 762\\
Crowther, P.A., Smith, L.J., Hillier, D.J., \& Schmutz, W.  1995,
  A\&A 293, 427\\
DeGioia-Eastwood, K., Throop, H., Walker, G., et al. 2001, ApJ 549, 578\\
Donati, J.-F., Wade, G.A., Babel, J., et al. 2001, MNRAS 326, 1265\\
Donati, J.-F., Babel, J., Harries, T.J.,  et al. 2002, MNRAS 333, 55\\
Dougherty, S.M., Williams, P.M., \& Pollaco, D.L. 2000, MNRAS 316, 143\\
Dougherty, S.M., \& Williams, P.M. 2000, MNRAS 319, 1005\\
Drew, J.E. 1989, ApJS 71, 267\\
Eichler, D., \& Usov, V. 1993, ApJ 402, 271\\
Evans, N.R., Seward, F.D., Kraus, M.I., et al. 2003, ApJ 589, 509\\
Garc\'{\i}a, B. Malaroda, S., Levato, H., et al. 1998, PASP, 110, 53 \\
Hog, E., Kuzmin, A., Bastian, U., et al. 1998, A\&A 335, L65 (Tycho-1)\\
Hog, E., Fabricius, C., Makarov, V.V., et al. 2000, A\&A 355, L27
  (Tycho-2)\\
Humphreys, R., \& McElroy, D.B. 1984, ApJ 284, 565 \\
van der Hucht, K.A. 2001, New Astronomy 45, 135 \\
van der Hucht, K.A., Setia Gunawan, D.Y.A., Williams, P.M., et al. 2001,
  in ``Interacting Winds from Massive Stars'', eds. A.F.J. Moffat \& N.
  St-Louis, ASP Conf. Ser. 165, p. 267\\
van der Hucht, K.A. 2002, Astrophysics \& Space Science 281, 199\\
Lamers, H.J.G.L.M., \& Leitherer, C. 1993, ApJ 412, 771\\
Langer, N., Hamann, W.-R., Lennon, M. et al. 1994, A\&A 290, 819\\
Leitherer, C., Chapman, J.M., \& Koribalski, B.S. 1995, ApJ 450, 289\\
Levato, H., Malaroda, S., Morrell, et al. 2000, PASP 112, 359 \\
Longair, M.S. 1997, High Energy Astrophysics, Vol. 2, Cambridge
  University Press, Cambridge\\
Ma\'{\i}z-Apell\'aniz, J., \& Walborn, N. R. 2002, in ``A Massive
  Star Odyssey: from Main Sequence to Supernova'', eds. K. A. van der
  Hucht, A. Herrero, \& C. Esteban, ASP Conf. Ser. 212, p. 560\\
Mason, B.D., Gies, D.R., Hartkopf, W.I., et al. 1998, ApJ 115, 821\\
Mason, B.D., Wycoff, G.L., Hartkopf, W.I., et al. 2001, AJ 122, 3466\\
Massey, P., \& Johnson, J. 1993, AJ 105, 980\\
Massey, P., \& Hunter, D.A. 1998, ApJ 493, 180\\
Miley, G.K. 1980, Ann. Rev. Astron. Astrophys. 18, 165\\
Miralles, M.P., Rodr\'{\i}guez, L.F., Tapia, M. et al. 1994,
  A\&A 282, 547\\
Morrell, N., Garc\'{\i}a, B., \& Levato, H. 1988, PASP 100, 1431\\
Panagia, N., \& Felli, M. 1975, A\&A 39, 1\\
Penny, L.R., Gies, D.R., Hartkopf, W.I., et al. 1993, PASP 105, 588 \\
Persi, P., Tapia, M., Rodr\'{\i}guez, L.F., et al. 1990, A\&A 240, 93\\
Prinja, R.K., Barlow, M.J., \& Howarth, I.D. 1990, ApJ 361, 607\\
Puls, J., Kudritzki, R.P., Herrero, A., et al. 1996, A\&A 305, 171\\
Rauw, G., Vreux, J.-M., \& Bohannan, B. 1999, ApJ 517, 416\\
Reed, B.C., \& Beatty, A.E. 1995, ApJS 97, 189\\
Reynolds, J. 1994, ``A Revised Flux Scale for the AT Compact Array'',
  ATNF Internal Report, AT/39.3/040\\
Scuderi, S., Panagia, N., Stanghellini, et al. 1998,
  A\&A 332, 251 \\
Setia Gunawan, D.Y.A., de Bruyn, A.G., van der Hucht, K.A., \& Williams, P.M.
  2000, A\&A 356, 676\\
Setia Gunawan, D.Y.A., van der Hucht, K.A.,
Williams, P.M., et al. 2001, A\&A 376, 460\\
Simon, K.P., Jonas, G., Kudritzki, R.P. et al. 1983, A\&A 125, 34\\
Tapia, M., Roth, M., V\'azquez, R.A. et al. 2003, MNRAS 339, 44\\
Taresch, G., Kudritzki, R.P., Hurwitz, M., et al. 1997, A\&A 321, 531\\
Usov, V. 1992, ApJ 389, 635\\
Vacca, W.D., Garmany, C.D., \& Schull, J.M. 1996, ApJ 460, 914  \\
V\'azquez, R.A., Baume, G., Feinstein, A., et al. 1996, A\&AS 116, 75\\
Vink, J.S., de Koter, A., \& Lamers, H.J.G.L.M. 2000, A\&A 362, 295 \\
Walborn, N.R. 1973, ApJ 179, 517 \\
Walborn, N.R. 1982, AJ 87, 1300\\
Walborn, N.R. 1992, ApJ 393, L13\\
Walborn, N.R. 2002, in ``A Massive Star Odyssey: from Main
  Sequence to Supernova'', eds. K. A. van der Hucht, A. Herrero, \&
  C. Esteban, ASP Conf. Ser. 212, p. 13\\
Walborn, N.R., Howarth, I.D., Lennon, D.J., et al. 2002, AJ 123, 2754\\
Waldron, W.L., Corcoran, M.F., Drake, S.A. \& Smale, A.P. 1998,
  ApJS 118, 217\\
Worley, C.E., \& Douglass, G.G. 1997, A\&AS 125, 523\\
White, R.L. 1985, ApJ 289, 698\\
Wright, A.E., \& Barlow, M.J. 1975, MNRAS 170, 41\\

\end{document}